\documentclass[3p]{elsarticle}

\usepackage{lineno,hyperref}
\modulolinenumbers[5]

\journal{High Energy Density Physics}

\begin{document}
	
	\begin{frontmatter}
		
		\title{Laser-driven generation of collimated quasi-monoenergetic proton beam using double-layer target with modulated interface}		
		\author{Martin~Matys$ ^{a,b,*}$}
				\cortext[mycorrespondingauthor]{Corresponding author}
				\ead{Martin.Matys@eli-beams.eu}
		\author{Katsunobu~Nishihara$ ^{a,c,d}$}
		\author{Mariana~Kecova$ ^{a}$}
		\author{Jan~Psikal$^{a,b}$}
		\author{Georg~Korn$^{a}$}
		\author{Sergei~V.~Bulanov$ ^{a,e,f}$}
		\address{$^{a}$ FZU - Institute of Physics of the Czech Academy of Sciences, ELI Beamlines, Na Slovance 2, 18221 Prague, Czech~Republic}
		\address{$^{b}$ Faculty of Nuclear Sciences and Physical Engineering, Czech Technical University in Prague, Brehova 7, Prague,~115~19,~Czech~Republic}
		\address{$^{c}$ Graduate School of Engineering, Osaka City University, Sugimoto-cho, Sumiyoshi, Osaka 558-8585, Japan}
		\address{$^{d}$ Institute of Laser Engineering, Osaka University, Suita, Osaka 565-0871, Japan}
		\address{$^{e}$ Kansai Photon Science Institute, National
			Institutes for Quantum and Radiological Science and Technology, 8-1-7~Umemidai, Kizugawa-shi, Kyoto 619-0215, Japan}
		\address{$^{f}$ Prokhorov Institute of General Physics of the Russian Academy of Sciences, Vavilova 38, Moscow,~119991,~Russia}
		

		\begin{abstract}
Usage of double-layer targets consisting of heavy and light material with modulated interface between them provides a way for laser-driven generation of collimated ion beams. With extensive 2D3V PIC simulations we show that this configuration may result in a development of a relativistic instability with Rayleigh-Taylor and Richtmyer-Meshkov like features. Initially small perturbations are amplified during the laser-target interaction leading to the formation of low-density plasma regions and high-density bunches between them, which are accelerated by the laser radiation pressure as whole compact structures. That results in collimated quasi-monoenergetic proton beam with high average energy. The properties of this proton beam such as its low emittance (one order of magnitude lower compared to that of conventional accelerators) and divergence are discussed. Results are compared with similar acceleration schemes such as double-layer target without corrugation and single-layer target.
		\end{abstract}
		
		\begin{keyword}
			ion acceleration\sep laser-driven\sep plasma\sep monoenergetic\sep instability \sep particle-in-cell simulation
		\end{keyword}
			
		\fntext[abbreviation]{Abbreviations: RTI: Rayleigh-Taylor instability, RMI: Richtmyer-Meshkov instability, SWI: Short-wavelength instability, HL: heavy-light, LH: light-heavy, HL-WO: HL without modulation, L2: light with the same thickness as HL,  HL-FF: HL with full-front laser pulse, L2-SM: L2 with surface modulation.}
	\end{frontmatter}
	
	
	\section{Introduction}
	
	Laser-driven ion accelerators has received a great deal of interest in last several decades as they are capable of sustaining relatively higher accelerating gradients than their conventional counterparts, and are currently able to accelerate protons to energies of 100 MeV \cite{Higginson2018_100MeV}. With the advent of multi-petawatt laser systems like ELI Beamlines (Czech Republic), APOLLON (France) or SEL (China) the laser pulses will soon reach intensities over $10^{23}\ \mathrm{W/cm^2}$ entering the acceleration regimes dominated by radiation pressure \cite{Esirkepov2004, SSB2016}, which promises proton/ion acceleration above energy of several GeV. 
	
	Studies of the high-intensity laser interaction with single-layer planar targets \cite{Pegoraro2007} shows the development of relativistic Rayleigh-Taylor \cite{Rayleigh1882,Taylor1950} like instability (RTI) leading to the formation of low-density regions and high-density ion bunches between them. The bunches exhibit quasi-monoenergetic behavior \cite{Pegoraro2007,Bulanov2010Unlim}. The instability can develop in a controlled way, when a corrugation is imprinted on the front surface of the target \cite{Echkina2010}. Bunches are then generated at the positions determined by the corrugation geometry. 
	
	The composite targets, consisting of planar heavy and light ion layers, have also been considered for generation of high quality ion beams \cite{Bulanov2002feas, Bulanov2002DL, Esirkepov2002, BULANOV2002_onco} required for various applications as hadron therapy
	\cite{Bulanov_2014_UspHadr} and nuclear fusion
	\cite{Roth2001, Atzeni2002}.
	When the corrugation is tailored on the interface between 
	two different layers, rather impulsive Richtmyer-Meshkov  \cite{Richtmyer1960,Meshkov1969} like instability (RMI) can develop. 
	
	These two instabilities belong to the same family group and are being thoroughly investigated \cite{Zhou2017,Palmer2012,Yang1994,Nishihara2010,Matsuoka2017,Zhou2019}, as they play important roles in various fields as the astrophysics (e.g., in the development of the filament structure of the Crab Nebula \cite{Hester2008}) and are affecting the creation of the hot spot in the inertial fusion \cite{Lindl1992}. The main differences between them are the dependence of the instability appearance on the duration of the driving force and the direction of the acceleration toward the interface \cite{Zhou2017}. The driving force of RTI is in principle continuous, while RMI is impulsive. RMI can occur when the acceleration is directed toward either side of the interface, whereas
	RTI can occur only for the direction from lighter to heavier media \cite{Zhou2017}. Moreover, in the case of heavy-light direction of acceleration, RMI exhibits characteristic phase inversion of the corrugated interface as was shown theoretically \cite{Yang1994,Velikovich1996,Wouchuk1996} and in the experiments \cite{Niederhaus2003}. Several theoretical models of RMI are known, including the exact linear solution \cite{Wouchuk1996}, asymptotic solution \cite{Fraley1986,Wouchuk1997}  and relativistic solution \cite{Mohseni2014}. Behavior of RMI can be also explained by the description of velocity shear induced at the corrugated interface  \cite{Nishihara2010,Matsuoka2017}.
	
	In this paper we present a positive effect of a controlled development of collisionless relativistic instability with RMI-like features on ion acceleration. The instability originates from the interaction of steep-front high-intensity, high-power laser pulse (with intensity of $10^{23}\  \mathrm{W/cm^2}$ and power of 80 PW) with a double-layer target having interface modulation. The assumed laser pulse is linearly polarized. This is in a direct contrast to several schemes for monoenergetic bunch generation, which have some similar features, but require circular polarisation. This includes, e.g., using of single-cycle laser pulses \cite{Zhou2016SC}; generation of self-organizing proton beam by stabilizing the central part of the foil, while letting the laser pulse propagate around it through unstable transparent wing regions \cite{Yan2009}; using of dual parabola target \cite{Liu2013}  or shaping the target in the transverse direction to match the laser intensity profile \cite{Chen2009}. The laser pulses assumed in these schemes (single-cycle and trapezoid) also inherently provide some sort of steep-front, required in our case, as the intensity rises from minimum to maximum in less (equal) then one laser period (trapezoid pulse profiles with rising ramps with the length of one and ten periods were compared in Ref. \cite{Yan2009}, longer ramp resulted in lower ion energy;  ramp with the length of five laser periods was used in Ref. \cite{Liu2013}). The circular polarisation is often proposed for the stabilisation of the foil by suppressing the instabilities \cite{Klimo2008PRST,Robinson2008}. Therefore, the linear polarisation is needed in our case to embrace the instability development.
  	
	Initially small perturbations at the interface are then amplified during the laser-target interaction, leading to the formation of low-density regions at the positions determined by the initial perturbation geometry and high-density plasma bunches between them. The bunches, with higher density than the density of the initial foil are then accelerated by the laser radiation pressure as whole compact structures. Moreover, the laser field propagating through the low-density regions enfolds the central plasma bunch, preventing from the perpendicular particle expansion. These behaviors result in the generation of quasi-monoenergetic, well-collimated ion beam with the average energy in the multi-GeV range and transverse emittance of one order of magnitude lower than that in the case of conventional accelerators. 
	The laser accelerated high-energy ion beams from composite targets may also find applications in material sciences and nuclear physics research \cite{Nishiuchi2015}.
	
	The paper is organized as follows. The simulation method and parameters are described in section \ref{Setup}. Section \ref{Results}, containing results,  is divided into four subsections. Firstly, the mechanisms of the development of the instability with RMI-like features and beam generation are described in \ref{R1}. Then the properties of the collimated quasi-monoenergetic proton beam such as its low emittance and divergence are discussed in \ref{R2}. The effects of different laser pulse polarisation and corrugation wavelength are studied in \ref{PolWL}. Lastly, our acceleration scheme is compared with schemes with different configurations such as using double-layer target without interface modulation, single-layer target and a case where full-front laser pulse is used instead of the steep-front pulse in \ref{R3}. After the Conclusions \ref{Conclusion}, the \ref{Appendix} follows, describing visualisation of our data in the form of figures, videos and a web-based application with virtual reality mode \cite{VBL}. The supplementary videos of time evolution of a selected simulation case are also included there.

	\section{Simulation method and parameters}\label{Setup}
	To demonstrate the ion acceleration scheme based on double-layer target with corrugated interface between two ion species and its advantages compared to targets without corrugations we performed 2D particle-in-cell simulations using the code EPOCH \cite{Arber2015}. The QED (quantum electrodynamics) module \cite{RIDGERS2014} resolving non-linear Compton scattering  is included in the simulations, since this phenomenon occurs in the assumed laser intensity range.
	
	In our case, linearly s-polarized (electric field is perpendicular to the plane of incidence) Gaussian laser pulse with a steep front incidents normally on the target.  The radiation wavelength is $ \lambda = 1\ \mathrm{\mu m}$ and the peak intensity is $ I_{\mathrm{max}} = 1.37\times10^{23}\  \mathrm{W/cm^2}$, thus yielding dimensionless amplitude $ a_0 = eE_0/m_e\omega c \approx 0.85\sqrt{I\left[10^{18}\mathrm{W\ cm^2}\right]\lambda^2\left[\mathrm{\mu m}\right]}  \approx 315 $. The critical plasma density is equal to $ n_c = \epsilon_0m_e \omega^2/e^2 \approx 1.115\times10^{21}\  \mathrm{cm^{-3}}$. Here, $ E_0 $ is the electric field amplitude, $ \epsilon_0 $ is permittivity of vacuum, $ \omega $ is laser angular frequency, $ m_e $ and $ e $ are electron mass and charge, respectively, and $ c $ is speed of light in vacuum.
	The amplitude $ a_0 $ can locally exceeds value of 500 in our simulations, due to the self-focusing. Therefore, implementing of radiation friction in QED regime \cite{Bulanov2015PPR} is required in our case. The laser beam width at the full width at half maximum (FWHM) is 10 $\lambda $ and beam duration at FWHM equals to 8 laser periods $ T $. The steep front is realized by filtering out the low-intensity part at the front of the laser pulse till 2.4 $ T $ (i.e., 30\% of FWHM) before the peak of the temporal Gaussian profile.   
	
	The laser profile can be produced by several methods, e.g., by using a thin overdense foil, so-called plasma-shutter \cite{Vshivkov1998,Reed2009,Palaniyappan2012,Wei2017,Matys2018ShutterEPS,Jirka2020} or it may occur due to the nonlinear evolution of the laser pulse propagating through an underdense plasma \cite{Bulanov1992Edge,Bulanov1993,Decker1996}. This approach may also improve spatio-temporal contrast of intense laser beam in possible future experiments by filtering a prepulse that accompanies the main pulse \cite{Mourou2006}. The generated steep front of the laser pulse can reduce the development of transverse short-wavelength  instabilities (hereinafter referred as SWI), as proposed in theory \cite{Pegoraro2007}. These instabilities cause the disruption of even initially planar foils during radiation pressure acceleration and are usually ascribed to RTI \cite{Pegoraro2007,Klimo2008PRST,Robinson2008} or electron-ion coupled instability \cite{Wan2016,Wan2018}. Reduction of SWI then enables the development of long-wavelength instabilities induced by the target geometry.
	
	The double-layer target consists of light and heavy ion layers. The light layer is made of solid hydrogen with electron density $ n_e =$ $5.36 \times 10^{22}\  \mathrm{cm}^{-3}$, i.e., 48 $ n_c $. It corresponds to targets already demonstrated in experiments \cite{Margarone2016} (with thickness down to $ 20~ \mathrm{\mu m} $). The heavy layer consists of corresponding cryogenic deuterium with the same electron and ion number density, but with two times heavier ion mass than in the case of the light layer. Therefore, the Atwood number $ A = (m_2\rho_2-m_1\rho_1)/(m_2\rho_2+m_1\rho_1) = \pm0.33 $, where $ m_{1,2},\ \rho_{1,2}  $ are the ion masses and densities at the front (1) and rear (2) layers. The sign depends on the direction of the acceleration, i.e., plus for the light-heavy (LH) case and minus for the heavy-light (HL) case. 
	
	The wavelength of the initial sinusoidal interface perturbation is $5\ \lambda$ and its amplitude is set to  $ 0.25\ \lambda$. The phase of the modulation is shifted by $ \pi $ between HL and LH cases in order to ensure the maximum number of proton particles around the  $ y$-axis. 
	Target thickness is set to 2 $ \lambda $ (1 $ \lambda $ per each layer), corresponding to the optimal thickness $ l $ for radiation pressure acceleration mechanism 
	\begin{equation}\label{eq.opt}
		\frac{l}{\lambda}=\frac{a_0n_c}{\pi n_e}.
	\end{equation}
	The radiation pressure acceleration mechanism starts to dominate over more traditional target normal
	sheath acceleration mechanism for these relatively low-density (but still overdense) targets, like cryogenic hydrogen, at even lower intensities \cite{Klimo2008PRST,Robinson2012,Psikal2018}. 
	
	The laser plasma interaction occurs in the simulation box with the size of $ 80\ \lambda\times 40\ \lambda$. The mesh has square cells. The size of the cells is set to $ 0.01\ \lambda $ to be shorter than the plasma skin depth $ c/\omega_{pe}\approx 0.02\  \lambda $, where $ \omega_{pe} $ is electron plasma frequency. Since the 3rd
	order b-spline shape of the quasi-particles and current smoothing are used in our simulations, it is ensured that numerical heating is strongly reduced even for the cells larger than the plasma Debye length \cite{Arber2015}. The simulation time step has been set by EPOCH code in order to satisfy CFL (Courant-Friedrichs-Lewy) condition \cite{Courant2010} to $ 6.7 \times 10^{-3}\ T $. Each cell inside the plasma slab initially contains 48 quasi-particle electrons and the same number of protons or deuterium ions, respectively. Temperatures of all particles are initialized to 5 keV to further reduce numerical heating. The particle solver begins to move the particles just a few time steps before the arrival of the laser
	pulse front to the target. Target is placed at the position $ x = 0 $, situated $ 10\ \lambda $ from the
	simulation box boundary in the direction of the laser propagation. The time instant, when the laser pulse front reaches the edge of the plasma is referred as $ t=0 $. The transverse size of the target is $ 40\ \lambda $, i.e., the target
	is reaching the simulation box boundaries at positions
	$ y = \pm\ 20\ \lambda$ where thermal boundary conditions for particles
	are applied.
	
	\section{Results}\label{Results}
	\subsection{Mechanisms of the beam generation}\label{R1}
\begin{figure}[ht]
		\begin{center}
			\includegraphics[width=\linewidth]{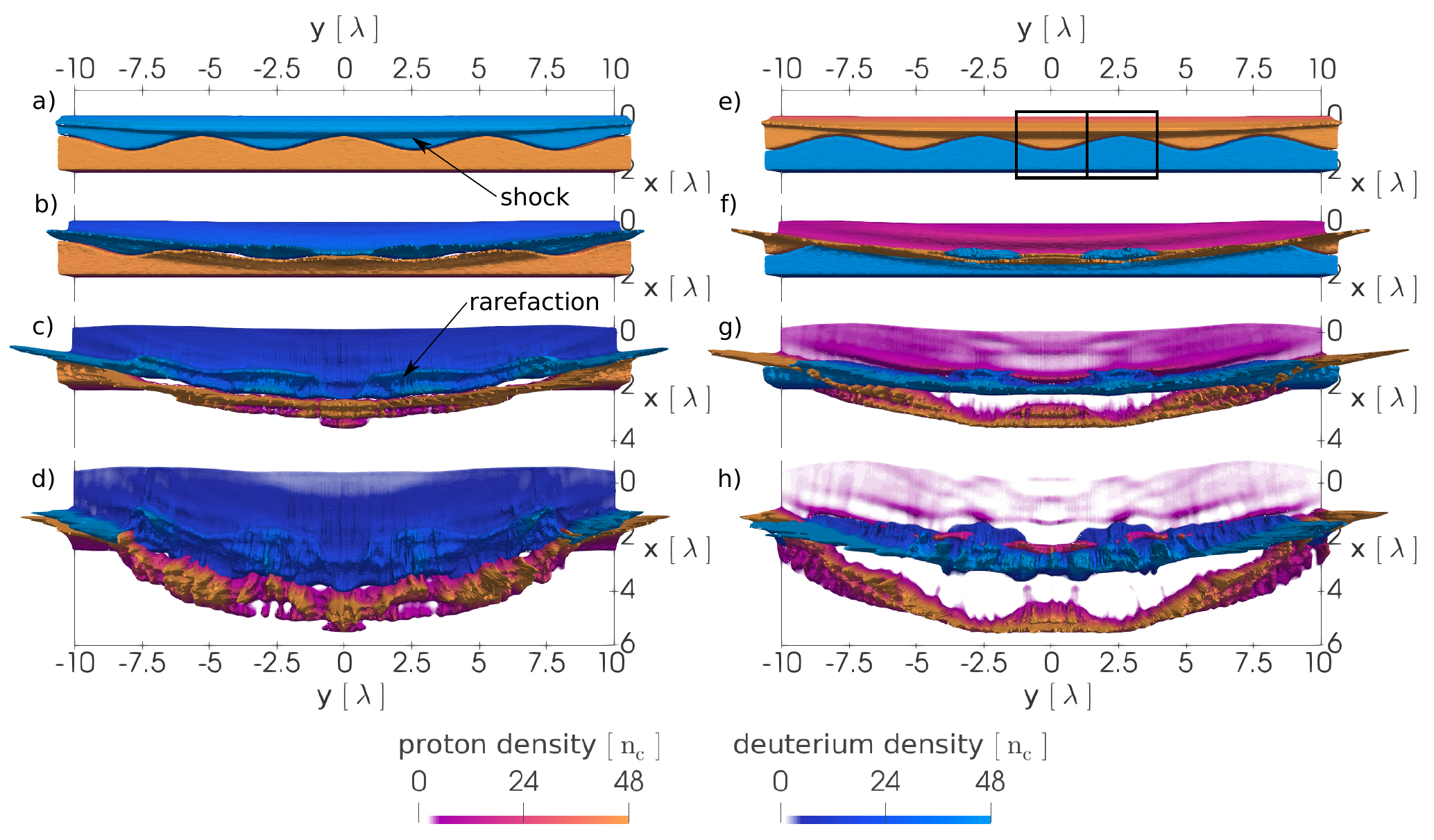}	
			\caption{\label{fig:RMImech}The development of the instability.  Presented cases: HL (left column)  and LH (right column). Blue and red scales represents deuterium and proton densities with maximum value set to the initial density. Full density is indicated by the vertical height. Time instants at rows:
				1.25 $ T $, 3 $ T $, 5.5 $ T $ and 8 $ T $.}
		\end{center}
	\end{figure}

	Despite its relatively low density, the target does not become fully relativistically transparent, as would be suggested by linear analysis  $n_e < \gamma n_{ec}$  \cite{Mora2001} (where $ \gamma=\sqrt{1+a_0^2/2}$ for linear polarisation), since electrons from the front layer are being pushed into the target by the ponderomotive force, piling up the initial electron density. Therefore, the radiation pressure can still be efficiently acting on the target \cite{Robinson2012}, driving a collisionless compression wave propagating toward the corrugated interface in the middle of the target (see Fig. \ref{fig:RMImech}-a and \ref{fig:RMImech}-e). For the sake of brevity and continuity with the currently established fluid RMI theory \cite{Zhou2017}, this shock-like jump discontinuity will be hereinafter referred as shock and its reflection in the HL case as rarefaction \cite{Yang1994}.  
	
	In the case of HL interaction, the shock reaches the interface at time $t = 1.25\ T $ (see Fig. \ref{fig:RMImech}-a). Therefore, the average shock speed is above 0.6 $ c $. The incident shock interacts with the corrugated interface (Fig. \ref{fig:RMImech}-b). The reflected rarefaction wave is observed in Fig. \ref{fig:RMImech}-c. Time evolution of phase inversion  of the corrugated interface (i.e., switching the positions of corrugation maxima and minima), characteristic for RMI in the HL case, can be observed comparing the areas at the left hand side of Fig. \ref{fig:RMImech}. Particularly, at the $ y $-positions $ \pm 2.5\  \lambda$ (inversion from corrugation minima in Fig. \ref{fig:RMImech}-a to maxima in Fig. \ref{fig:RMImech}-d) and at the $ y $-positions $ 0\ \lambda$ and $\pm 5\ \lambda$ (inversion from corrugation maxima to minima). The phase inversion results in the stretching of the proton layer as regions of initial corrugation minima stay behind the regions of initial maxima (Fig. \ref{fig:RMImech}-d). It subsequently creates low-density regions between them at the positions, where the initial amplitude of the corrugation was zero ($ \pm 1.25\ \lambda $ and $ \pm 3.75$ $ \lambda $). 
	
	Different situation occurs in the case of LH interaction. After the shock hits the interface, the proton layer enters into the deuterium one and eventually propagates through it, as can be seen in the time evolution at the right hand side of Fig. \ref{fig:RMImech}. Now the phase on the remaining interface (at $ x $-position around $ 2\ \lambda $ in Fig. \ref{fig:RMImech}-g and -h ) is kept (maxima stay at the $ y $-positions $ \pm 2.5\  \lambda$ and minimum at $0\ \lambda$) as predicted by the RMI theory.  Deuterium layer becomes (relativistically) transparent to the incident laser pulse and the radiation pressure is acting on the detached proton layer around $ x $-positions $3\ \lambda$ (Fig. \ref{fig:RMImech}-g) and $5\ \lambda$ (Fig. \ref{fig:RMImech}-h). The detached proton layer then undergoes phase inversion. However, the driving mechanism is essentially different from that in the HL case and can be explained as follows. The momenta delivered to the particles by the laser pulse in both rectangles presented in Fig. \ref{fig:RMImech}-e are approximately the same (neglecting effects of Gaussian shape of the pulse around axis). However, the number of protons in the central rectangle (around $ y $-position $ 0\ \lambda $) and lateral rectangle differ, i.e., less particles receive the same amount of momenta from the laser pulse in the case of lateral rectangle. Therefore, they can reach higher energies, propagate with higher velocity and eventually overtake the particles initially located inside the central rectangle.
	
\begin{figure}[ht]
			\begin{center}
				\includegraphics[width=\linewidth]{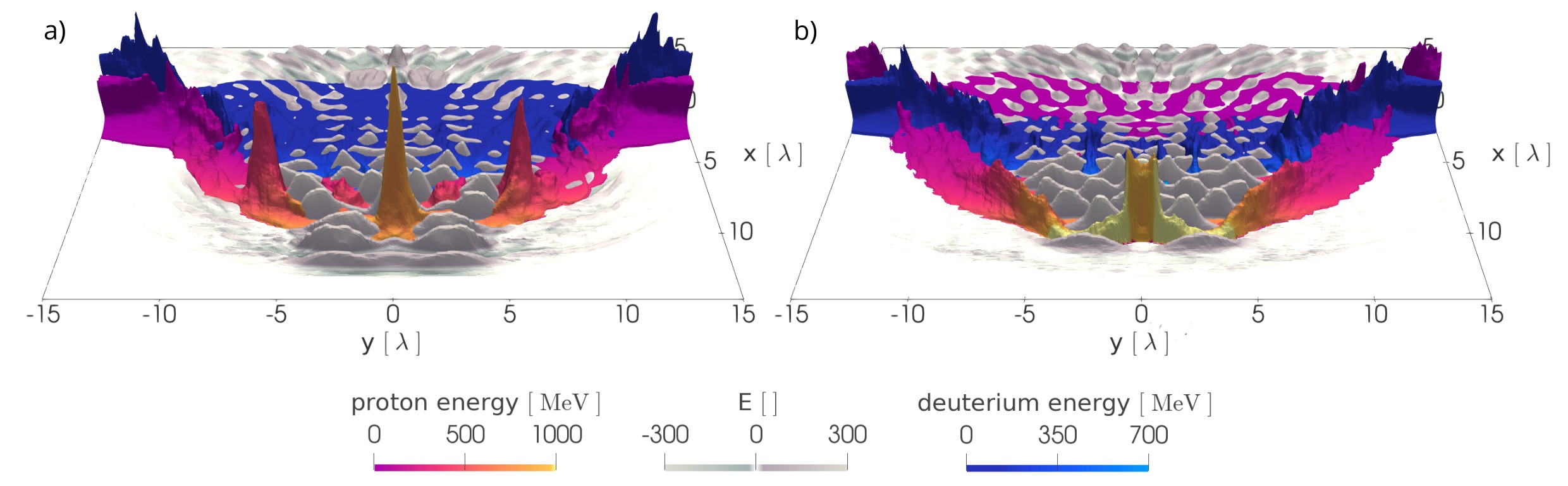}
				
				\caption{\label{fig:HustEn}Spatial distributions at time $ t = 14\ T $. The ion local mean energy, represented by red and blue color scales, their density, represented by the vertical height and the laser pulse electric field in the z-direction (represented by both the vertical height and by the grey scale). Simulated cases: a) HL, b) LH.}
			\end{center}
		\end{figure}

	To highlight the positions of the high-energy particles at later time ($ t = 14\ T $), the energies of the particles are being displayed as colors in Fig. \ref{fig:HustEn} instead of density as in Fig. \ref{fig:RMImech}. The laser pulse (depicted with grey color) is also included in this figure.
		  
	In the HL case (Fig. \ref{fig:HustEn}-a), the protons on the regions around initial maxima ($ 0\ \lambda$ and $\pm 5\ \lambda$) are accelerated to higher energies than around initial minima ($\pm 2.5\ \lambda$), which stay behind. Central and lateral bunches are being developed at the positions of initial maxima. The laser pulse propagates through the (relativistically) transparent low-density regions. The propagating electric field creates areas of high ponderomotive potential. Electrons are then pushed to the area of lower ponderomotive potential (i.e., lower electric field $ E $) around the axis by the ponderomotive force, subsequently reducing the perpendicular movement of the ions.  
	This field is not enfolding the lateral bunches (around $ y $-positions $\pm 5\ \lambda$) which then rapidly dissipate in time and only central bunch is present at time $ t = 47\ T $ as will be shown in Fig. \ref{fig:DiffMechN_D}-a. Supplementary videos of time evolution of this case can be found in \ref{Appendix}. 
	
	In the LH case (Fig. \ref{fig:HustEn}-b) the central bunch reaches lower energies then lateral areas (around $ y $-positions $\pm 2.5\ \lambda$), corresponding to the previous discussion about rectangles in Fig. \ref{fig:RMImech}-e. This situation is opposite to the HL case. The detached proton layer can confine the laser pulse and can be accelerated as a bubble \cite{Pegoraro2007}, till the low-density regions appears in the lateral areas (initial maxima at the y-positions $\pm 2.5\ \lambda$). Therefore, the maximal reached energy is still slightly higher at time $t = 14\ T $ in the LH case (light yellow color in Fig. \ref{fig:HustEn}-b) than in the HL case  (Fig. \ref{fig:HustEn}-a). However, due to the occurrence of the low-density regions at the positions of initial zeros of the corrugation (HL case) instead of maxima  (LH case), more narrow bunch with higher density and also more narrow enfolding field develop in the HL case. This behavior is crucial for long term acceleration as will be shown later.
	
	\subsection{Beam energy and quality}\label{R2}
\begin{figure}[ht]
			\begin{center}
				\includegraphics[width=\linewidth]{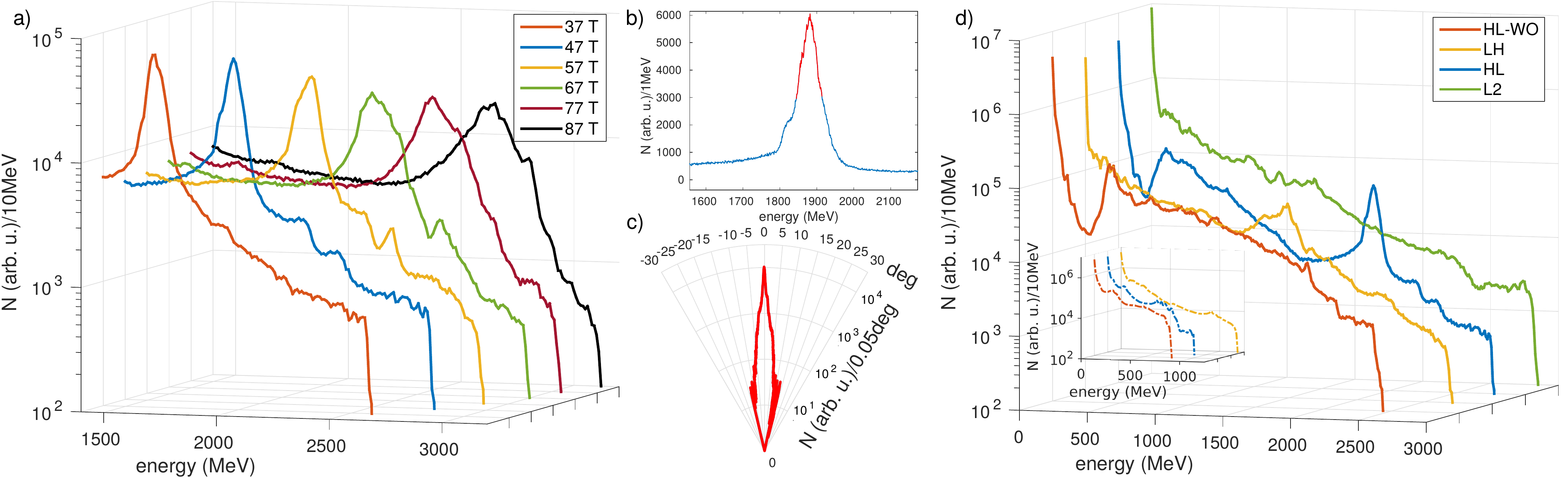}	
				\caption{\label{fig:Spektravse}Properties of the ion beam. a) Time evolution of the tail of the proton energy spectra in the HL case, b) proton energy spectra of the HL case at time $ t = 47\ T $, with highlighted FWHM section used for c) angular distribution, d) proton energy spectra (corresponding deuteron energy spectra in inset) for various targets (see details in the text) at time $ t = 47\ T $.}
			\end{center}
		\end{figure}

	In the HL case, well collimated quasi-monoenergetic proton bunch is developed as can be seen in Fig. \ref{fig:Spektravse}-a, where time evolution of proton energies is shown.
	
	 Although the energy spread is increasing after time $ t = 47\ T $, the bunch structure is still kept till the end of the simulation. Moreover, the average bunch energy is gradually shifting closer to the maximum energy during time. The average beam energy at time $ t = 47\ T $ reaches 1882 MeV and bandwidth (at FWHM) is 69 MeV (Fig. \ref{fig:Spektravse}-b). Therefore, the energy spread is about 3.7 \%. Angular distribution of the ions in the beam (red part of Fig. \ref{fig:Spektravse}-b) is shown in Fig. \ref{fig:Spektravse}-c. This graph implies low angular spread of $2\theta = 0.65^\circ$ (at FWHM). Therefore, the solid angle is $ \Omega = 2\pi\left(1-\cos\left(\theta\right)\right) = 0.1\ \mathrm{mrad} $. The normalized rms transverse emittance of these particles is $ \epsilon_{rms} = \sqrt{\langle y^2\rangle\langle p^2_y\rangle - \langle yp_y \rangle^2 }/m_pc = 0.046\ \mathrm{mm\cdot mrad}$. Where $ m_p $ and $p_y$ are proton mass and momentum in the (transverse) y-direction. $\epsilon_{rms}$ is proportional to the area of the ellipse containing particles in the phase space ($ y $-$ p_y $).
	Referring definition and further discussion can be found in Ref. \cite{Emittance2}. 
	This emittance is one order of magnitude lower than in the case of conventional proton accelerators \cite{Zhang2018}, but still one order of magnitude higher than the emittance reported in  Ref. \cite{Cowan2004} for much lower energy range of protons up to 10 MeV.
	
	The transverse emittance can be also defined in real space via the beam divergence 
	($ \Theta_{div}=\sqrt{\sum_{i=1}^{N}\frac{\left(\Theta_i-\left<\Theta\right>\right)^2}{N}} $) as in \cite{Gu2014} 
	
	\begin{equation}
		\epsilon_y=\frac{4}{N}\sqrt{\sum_{i=1}^{N}\left(y_i-\left<y\right>\right)^2}\times \sqrt{\sum_{i=1}^{N}\left(\Theta_i-\left<\Theta\right>\right)^2}.
	\end{equation}
	That yields the values for protons at the FWHM of the bunch (Fig \ref{fig:Spektravse}-b): $ \Theta_{div} = 0.038\ \mathrm{rad}$  and  $ \epsilon_y = 0.218 \ \mathrm{mm\cdot mrad} $. Assuming the whole high-energy proton beam (from 1748 MeV, i.e., the beginning of the bunch waist, to the maximum beam energy of 2770 MeV) the values rise to $ \Theta_{div} = 0.051$ and $ \epsilon_y = 0.372 \ \mathrm{mm\cdot mrad}$. This transverse emittance is still two order of magnitude lower compared to the one in Ref.  \cite{Gu2014}, in which protons reached similar energies to our case (1.67 GeV) in the simulation with underdense hydrogen target.
	
	Approximately  $ 0.32\times10^6 $ quasi-particles have energy within the assumed energy range, which means about 6.7\%  of $4.8\times10^6 $ quasi-particles initially located inside the laser spot area (i.e., $ \pm 5\ \lambda $). That yields $ 1.43\times10^{12}$ real particles and charge of 229 nC, assuming a projection of the initial 2D flat target of $ 2\times40\  \mathrm{\mu m^2}$ into a 3D cuboid of $ 2\times40\times40\  \mathrm{\mu m^3}$. Around 40.4 \% of the laser pulse energy is converted into the particles propagating in the forward direction  (protons: 28.7 \%, deuterons: 6.1 \%, electrons: 5.6 \%). Around 3.4 \% of the laser pulse energy is converted into the proton particles in the energy range of $ 1882\ \pm 69 $ MeV (corresponding to the FWHM red part of Fig. \ref{fig:Spektravse}-b).
	
	\subsection{Effects of different laser pulse polarisation and corrugation wavelength}\label{PolWL}
	
	Linear s-polarisation and single corrugation wavelength was used for the simulations so far. Other polarisations and corrugation wavelengths are investigated in this section, as they can affect laser-plasma interaction. The blue color line is used in graphs throughout the paper to highlight the HL case with the default parameters described in the section \ref{Setup}. 
	
	The employment of the circular polarisation (C-pol) instead of linear one can mitigate the instability development, inhibits a strong electron heating and stabilize the foil as was shown, e.g., in Refs. \cite{Klimo2008PRST,Robinson2008,Yan2009}. In the case of linear polarisation, the differences between p-polarisation (P-pol) and s-polarisation (S-pol) may become especially significant in the 2D geometry, as the laser polarisation is in or out of the simulation plane, respectively. 
	
\begin{figure}[ht]
		\begin{center}
			\includegraphics[width=\linewidth]{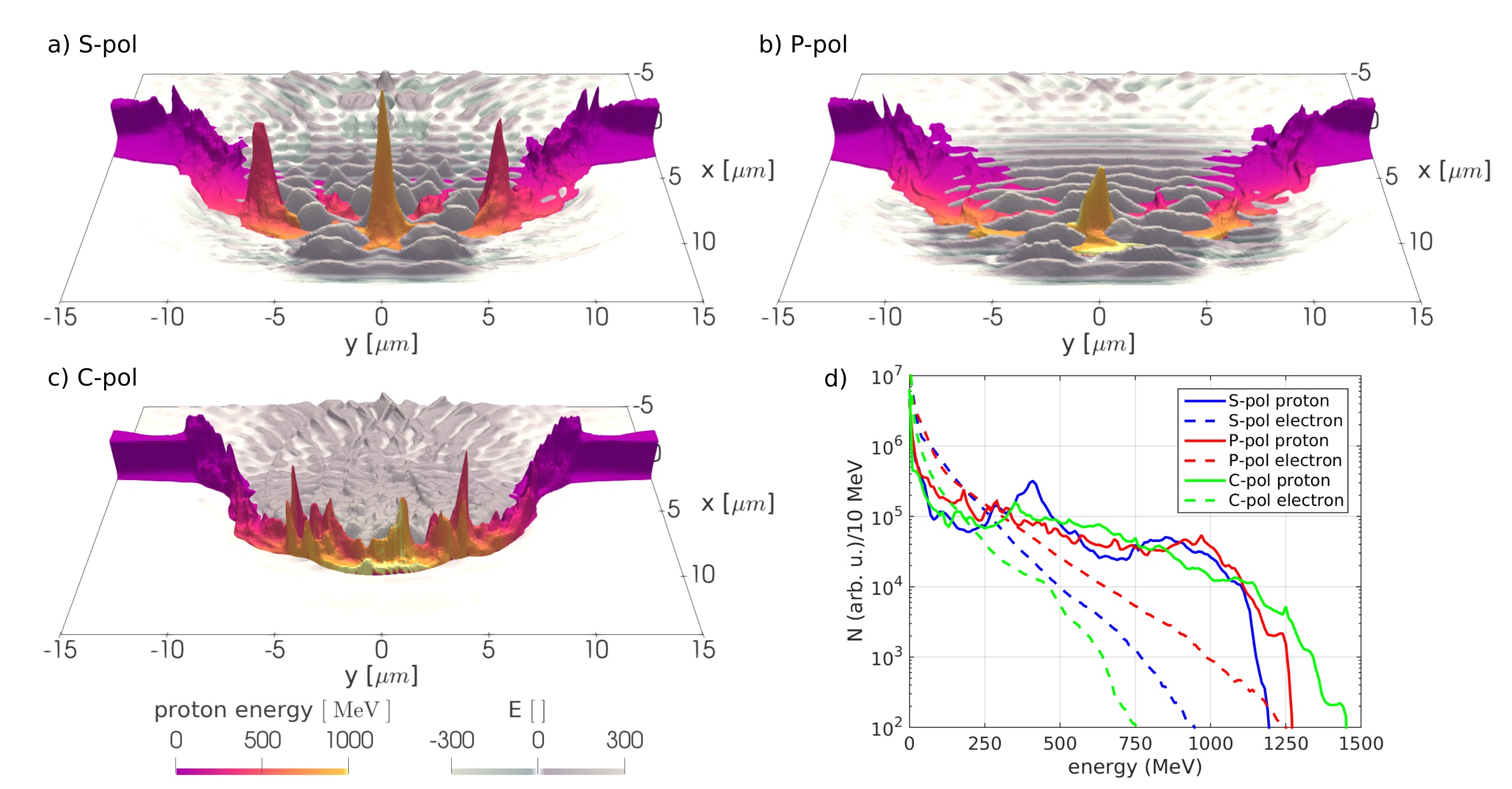}
			
			\caption{\label{fig:Pol14} Spatial distributions and energy spectra at time $ t = 14\ T $ for different laser pulse polarisations: a-c) The proton local mean energy, represented by red color scales, its density, represented by the vertical height and the laser pulse electric field in the polarisation dependent direction (represented by both the vertical height and by the grey scale), d) proton and electron energy spectra for the corresponding cases.}
			
		\end{center} 
	\end{figure}

	The effects of the laser pulse polarisations on the early laser-plasma interactions (at time $ t = 14\ T $) of our case are visualised in the terms of proton density, proton local mean energy and laser pulse electric field in Fig. \ref{fig:Pol14}. 
	
	In the case of S-pol (our default case) the instability developed, resulting into the generation of three bunches (with distinctive central one), where the radiation pressure is taking place. Laser pulse than can propagate only through the low-density regions between them (Fig. \ref{fig:Pol14}-a). The instability developed also in the case of P-pol (Fig. \ref{fig:Pol14}-b). However, the lateral bunches already dissipated and most of the foil is becoming (relativistically) transparent to the incoming laser pulse. Only the central bunch still holds, but in a significantly smaller form compared to the S-pol case.
	This behavior can be explained by artificially greater electron heating in the simulation plane in the P-pol case, while the S-pol case provides more isochoric heating into all 3 spacial directions as was demonstrated in Refs. \cite{Liu2013pol,Stark2017}. Moreover, the target will become (relativistically) transparent earlier in the P-pol case \cite{Stark2017}, which affects the generation of low-density regions and then of the bunch itself as was described in our simulations in the section \ref{R1}. The extra heating of electrons in the P-pol case is visible in the electron energy spectra in Fig. \ref{fig:Pol14}-d as the electrons are accelerated to higher energy with higher temperature (which can be inferred from the flatter slope of the curve compared to the S-pol case).
	
	On the contrary the instability is mitigated in the case of C-pol (Fig. \ref{fig:Pol14}-c). The laser pulse is confined by the bubble shaped foil and radiation pressure is properly acting on the whole area of the laser focal spot. The electron heating is reduced and protons are accelerated to higher energy than both S-pol and P-pol cases by the radiation pressure as can be seen in Fig. \ref{fig:Pol14}-d. However, due to the lack of instability development no distinctive bunch structure is found in the proton energy spectra for the C-pol case. 
	
\begin{figure}[ht]
		\begin{center}
			\includegraphics[width=\linewidth]{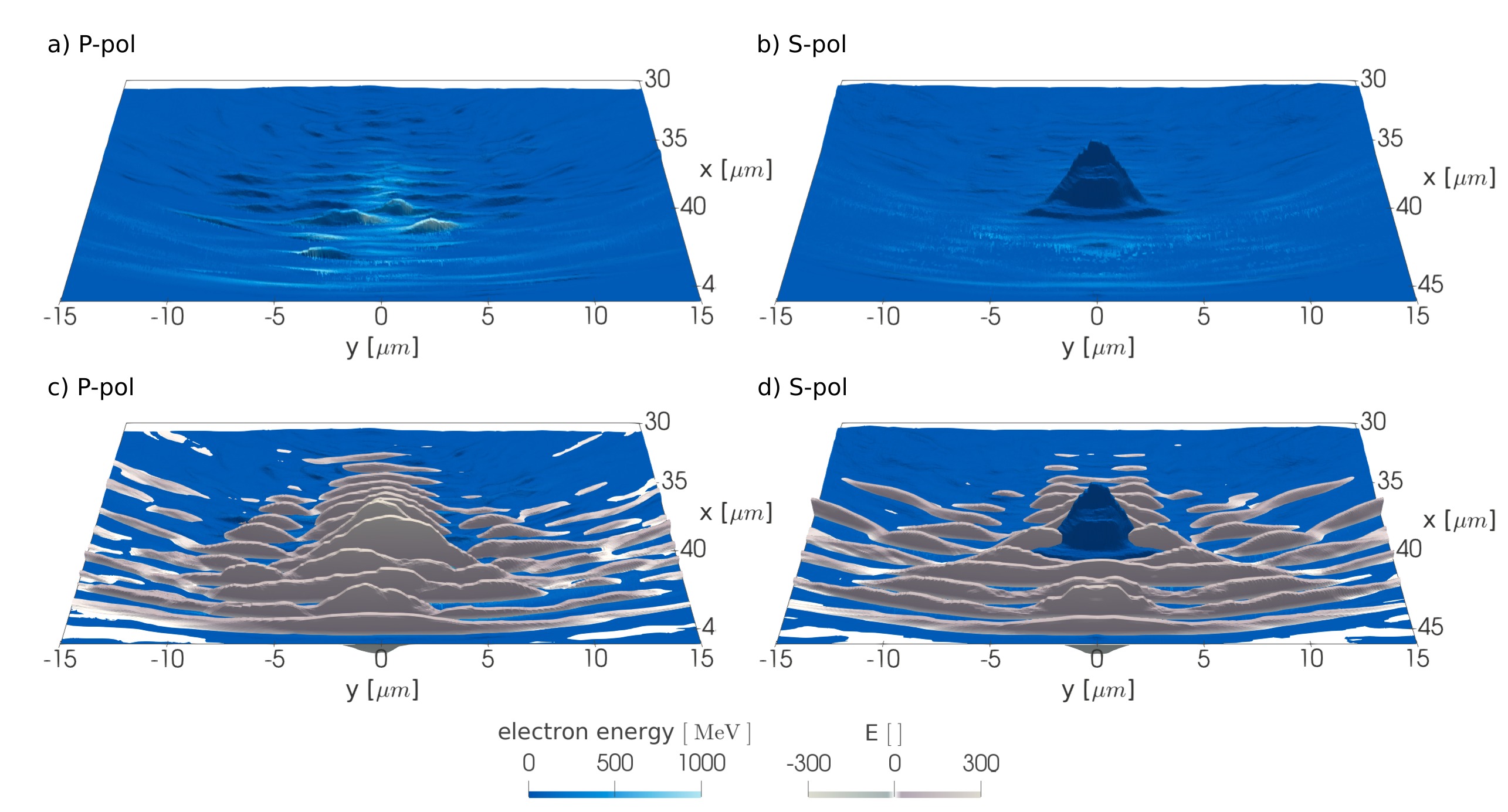}
			
			\caption{\label{fig:Pol_el} Spatial distributions at time $ t = 47\ T $ for different laser pulse polarisations: electron local mean energy, represented by blue color scales, its density, represented by the vertical height and the laser pulse electric field in the corresponding direction (represented by both the vertical height and by the grey scale).}
			
		\end{center} 
	\end{figure}

	At the later time $ t = 47\ T $ the P-pol case becomes mostly (relativistically) transparent for the laser pulse with only a few structures in the electron density as can be seen in Fig. \ref{fig:Pol_el}-a and \ref{fig:Pol_el}-c. On the contrary, the distinctive bunch structure is kept in the electron density in the S-pol case (Fig. \ref{fig:Pol_el}-b). The radiation pressure acceleration of the bunch is ongoing, while the area around bunch is (relativistically) transparent as can be seen in Fig. \ref{fig:Pol_el}-d. Note that the energy of electrons is relatively low inside the bunch (which means its thermal expansion is reduced) and is significantly higher in the transparent plasma where the electrons are heated and are oscillating directly in the laser field.
	
\begin{figure}[ht]
		\begin{center}
			\includegraphics[width=\linewidth]{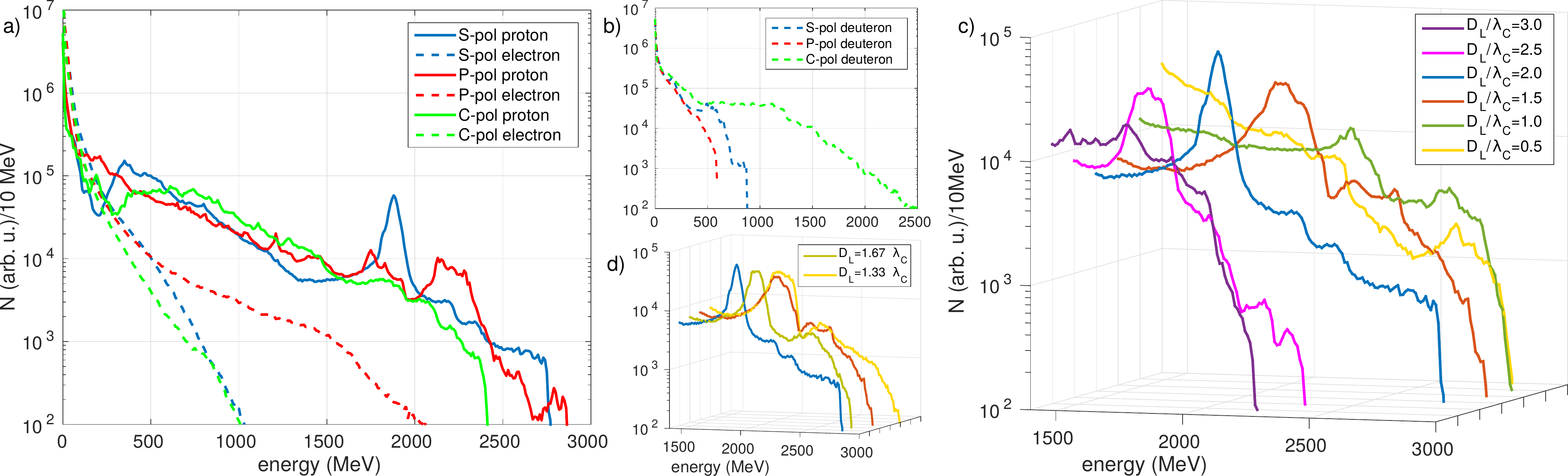}
			
			\caption{\label{fig:RefSpektra} Energy spectra at time $ t = 47\ T $ for various parameters. a) Proton and electron energy spectra for simulation study cases with different laser pulse polarisations, b) corresponding deuteron energy spectra, c-d) the tail of the proton energy spectra of various corrugation wavelengths $ \lambda_C $ for the focal spot diameter $ D_L=10\ \mathrm{\mu m} $.}
			
		\end{center} 
	\end{figure}

	The extraordinary electron heating in the P-pol case becomes even more apparent in the electron spectra at time $ t = 47\ T $ (Fig. \ref{fig:RefSpektra}-a) with reaching of the transparency regime. On the contrary, the proton energies are only slightly higher in the P-pol case, which corresponds to the discussion in Ref. \cite{Liu2013pol} for the transparency regime. The stabilisation effect of circular polarisation is also visible in the deuteron spectra in Fig. \ref{fig:RefSpektra}-b as the deuterons are accelerated to significantly higher energies than in the cases with linear polarisations. The distinctive bunch structure is visible only in the proton spectra for linear s-polarisation, which is then employed in the simulations hereinafter. The dependence on linear polarisation clearly distinguishes our scheme from similar ones described in the introduction \cite{Zhou2016SC,Yan2009,Liu2013,Chen2009} as the circular polarisation plays a crucial role in them.
	
\begin{figure}[ht]
		\begin{center}
			\includegraphics[width=\linewidth]{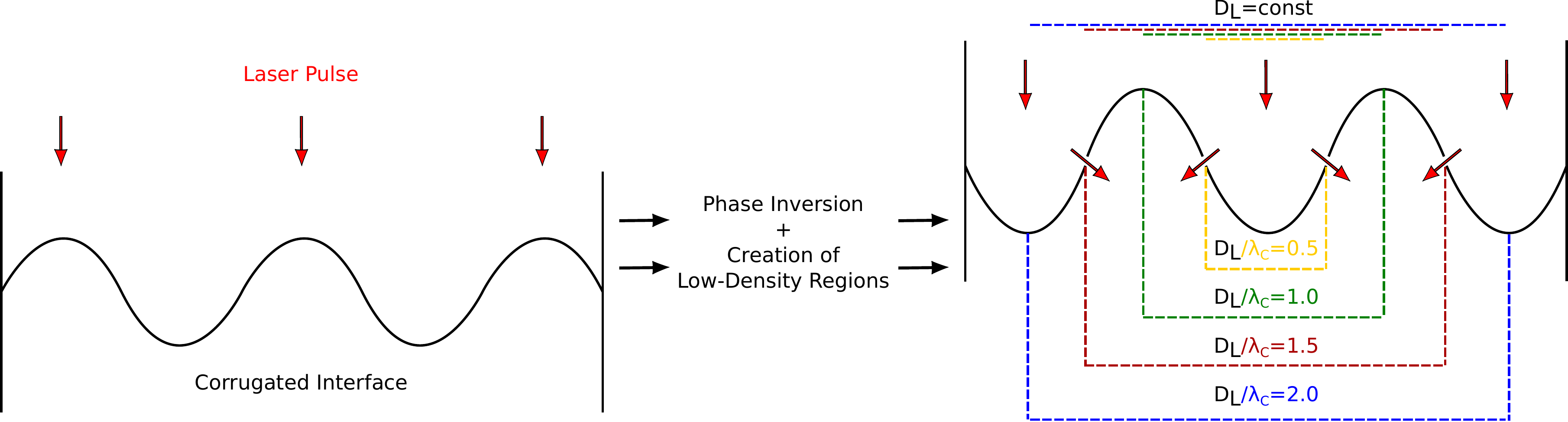}
			
			\caption{\label{fig:Schema} Scheme of the different geometric interaction for various corrugation wavelengths $ \lambda_C $.}
			
		\end{center} 
	\end{figure}

	Another parameter significantly affecting the ion acceleration in our scheme is the wavelength of the interface corrugation $ \lambda_C $. As was demonstrated in section \ref{R1}, the bunches are generated at the positions of corrugation maxima and are dependent on the development of low-density regions, located at the corrugation zeros. Therefore, a change of the corrugation wavelength may significantly modify the geometry of the system with a constant finite focal spot of diameter $ D_L $. The increase of the ratio $ D_L/\lambda_C $ by 0.5 then corresponds to the addition of another two corrugation points (maxima / minima / zeros) into the system as can be seen in the scheme in Fig. \ref{fig:Schema}.

\begin{figure}[ht]
		\begin{center}
			\includegraphics[width=\linewidth]{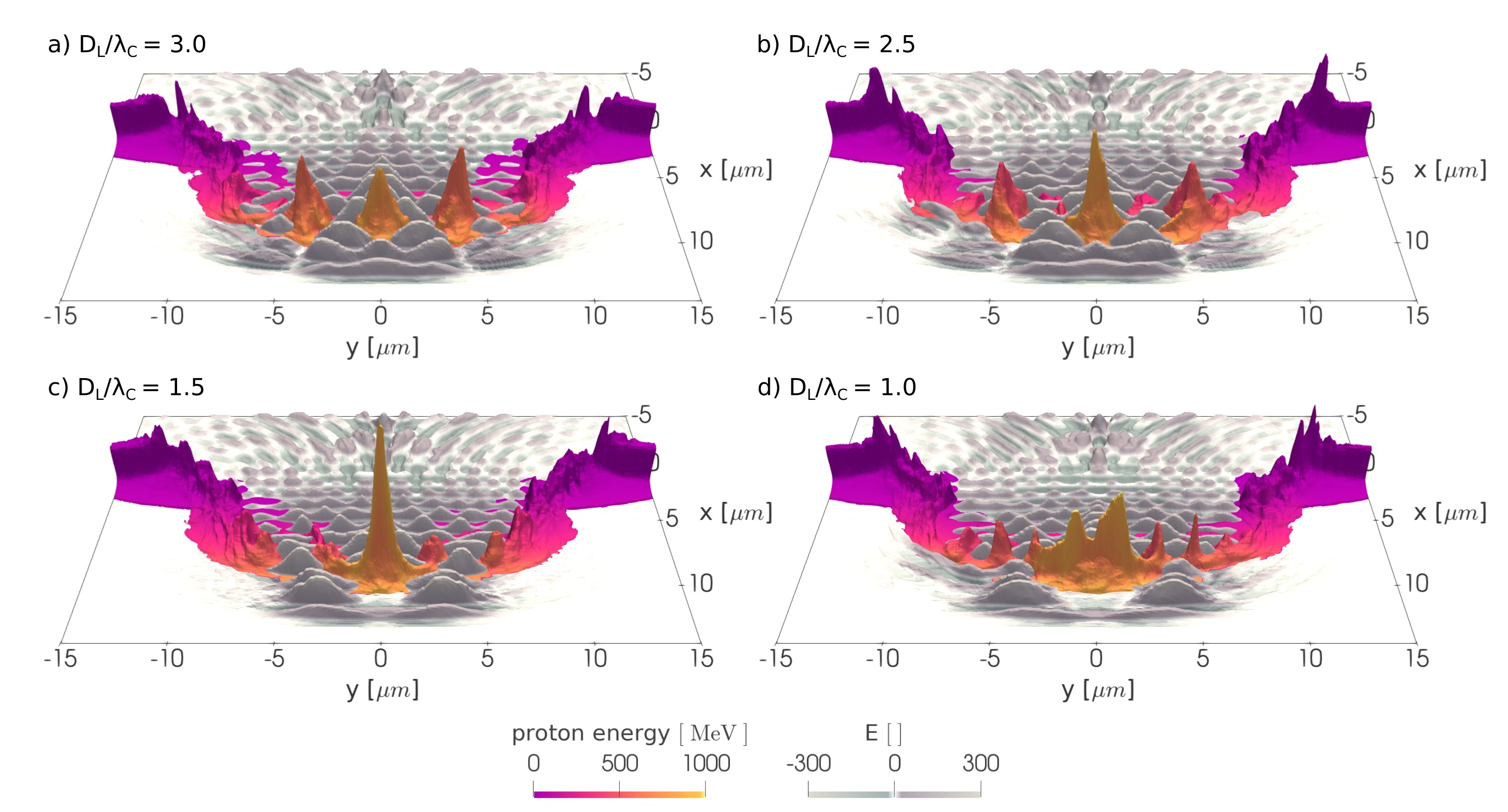}
			
			\caption{\label{fig:WL} Spatial distributions at time $ t = 14\ T $ for different corrugation wavelengths. The proton local mean energy is represented by red color scales, proton densities are represented by vertical height. The laser pulse electric field (in the z-direction) is represented by both the vertical height and by grey scale.}
			
		\end{center} 
	\end{figure}

	The HL simulation described in \ref{R1} corresponds to the ratio $ D_L/\lambda_C = 2.0 $ with 4 corrugation zeros. The low-density regions developed at the zeros further from the central axis provides a channels converging the laser pulse toward the central axis as can be seen in Fig. \ref{fig:HustEn}-a and in the scheme in Fig. \ref{fig:Schema}. Therefore, an efficient bunch should not be generated for the ratio $ D_L/\lambda_C< 1 $, where this phenomena cannot occur. For the same reason the bunch structure should significantly deteriorate for the ratio $ D_L/\lambda_C> 2.5 $ as another diverging channel is being applied. This idea corresponds to proton spectra shown in Fig. \ref{fig:RefSpektra}-c and is also valid for ratios $ D_L/\lambda_C $ which are not multiples of 0.5 as shown in Fig. \ref{fig:RefSpektra}-d. The average bunch energy and maximum proton energy in simulation is rising with the corrugation wavelength (decreasing with $ D_L/\lambda_C $) till $ D_L/\lambda_C>1 $. The lowest energy spread was achieved in the case of $ D_L/\lambda_C = 2 $ (i.e. the  HL case from the section \ref{R2}). The energy spread is rising with both increasing and decreasing corrugation wavelength. Two phenomena play roles in this case. The number of particles in the central bunch between the positions $ -1/4 \lambda_C $ and $ +1/4 \lambda_C $  (zeros closer to the central axis) is increasing with $ \lambda_C $ as well as the enfolding field is broadened (see Fig. \ref{fig:WL}). There, the proton local mean energy and density for cases $ D_L/\lambda_C = 3.0 $, $ D_L/\lambda_C = 2.5 $, $ D_L/\lambda_C = 1.5 $ and $ D_L/\lambda_C = 1.0 $ corresponding to Fig. \ref{fig:HustEn}-a at time $ t = 14 $ T are shown. On one hand, if the corrugation wavelength is too small as shown in Fig. \ref{fig:WL}-a, more bunches of similar low density are actively developed. On the other hand, if the corrugation wavelength is too large, the enfolding field is too broad and the central bunch spreads in space as shown in Fig. \ref{fig:WL}-d. Therefore, an optimum corrugation wavelength is around $ D_L/\lambda_C = 2.0$ which is then used for comparison with similar types of target in the following section.

	\subsection{Comparison with similar types of targets}\label{R3}
\begin{figure}[ht]
			\begin{center}
				\includegraphics[width=\linewidth]{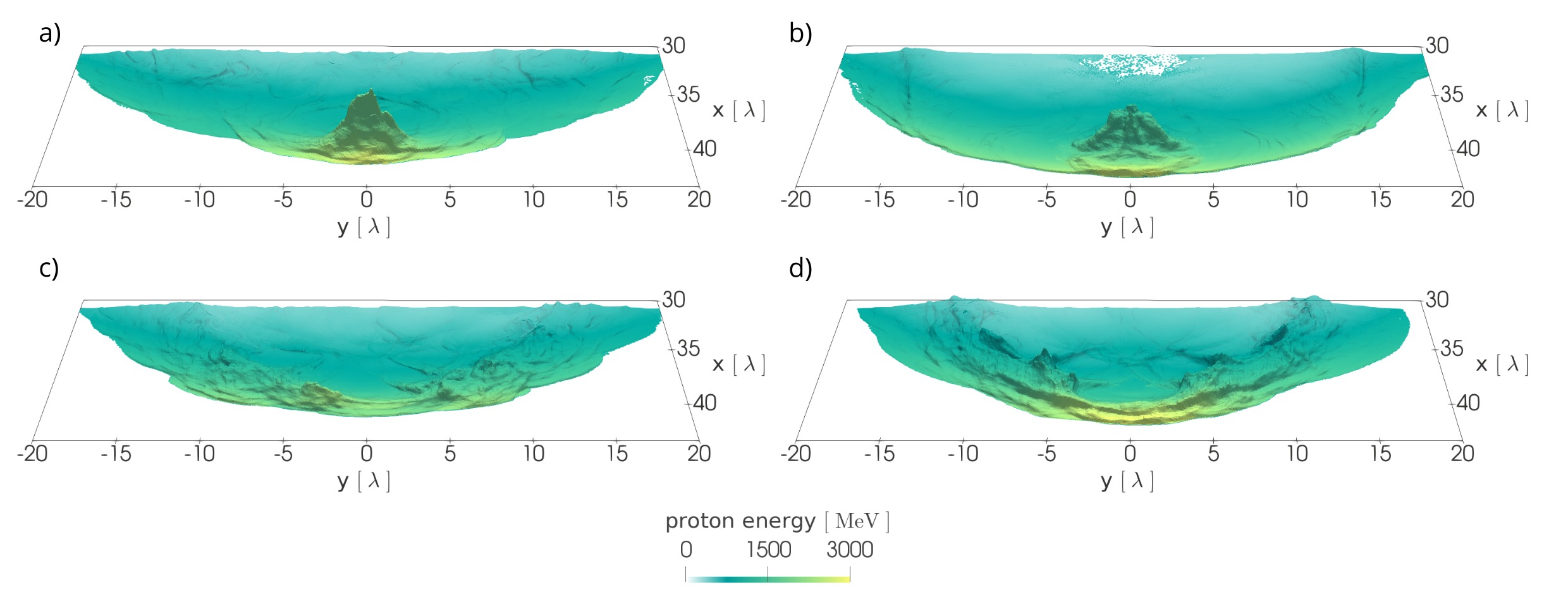}
				
				\caption{\label{fig:DiffMechN_D}Spatial distributions at time $ t = 47\ T $.  The proton density is represented by the vertical height and proton energy is represented by the blue to yellow scale. The simulated cases are: a) HL, b) LH, c) HL-WO, d) L2 (see parameters of all the targets in the text). }
				
			\end{center} 
		\end{figure}

	To point out the effects of double-layer target and modulation on its interface, another two simulations are performed. Namely, double-layer deuterium-hydrogen target without modulation, referred as HL-WO, and pure hydrogen target of the same thickness ($ 2 \lambda $), referred as L2. The reference time was chosen to be $ t = 47\ T $ as in the previous discussion. Proton energy spectra of all the simulated cases are shown in Fig. \ref{fig:Spektravse}-d and corresponding proton density and energy spatial distributions are presented in Fig. \ref{fig:DiffMechN_D}.
	
		 In the HL and LH cases (with the interface modulation) a bunch structure is created in the spatial plot (Fig. \ref{fig:DiffMechN_D}-a and -b) and corresponding quasi-monoenergetic peak is developed in the proton energy spectra (Fig. \ref{fig:Spektravse}-d).
	The trend of a more narrow bunch in the HL case compared with the LH case, initialized by different positions of the low-density regions of the foil, also continues in the later time.  This results into more narrow peak in the energy spectra with higher average energy in the HL case (1890 MeV compared to 1490 MeV)  in Fig. \ref{fig:Spektravse}-d. Maximal reached energy is also slightly higher in the HL case (2770 MeV compared to 2700 MeV). Proton bunch is more shifted in space towards the front of the overall proton cloud and less spread in the $ x $-direction in the HL case (compare Figs \ref{fig:DiffMechN_D}-a and \ref{fig:DiffMechN_D}-b).  
	
	On the contrary, HL-WO and L2 cases develop a bubble structure (Fig. \ref{fig:DiffMechN_D}-c and \ref{fig:DiffMechN_D}-d), with no significant peaks in the energy spectrum (Fig. \ref{fig:Spektravse}-d). The maximum proton energy of the simulated cases is reached in the L2 case (2850 MeV). This case is the best match for the radiation pressure acceleration regime described by Eq. \ref{eq.opt} assuming single layer target with no induced instability. The symmetric bubble structure holds till the reference time and proportionally small amount of protons near the axis are accelerated to the highest maximal energies (Fig. \ref{fig:DiffMechN_D}-d). Oppositely, the bubble shape is getting distorted in the HL-WO case (Fig. \ref{fig:DiffMechN_D}-c), due to the multi-ion-species effects, resulting into significantly lower maximum energy (2430 MeV).
	
	At the same reference time deuterium ion reached significantly lower maximal energies than protons in our simulations (see inset in Fig.  \ref{fig:Spektravse}-d). Their energies are lower than 1200 MeV (i.e., 600 MeV per nucleon)  and the highest energy is reached in the LH case. It corresponds to the observation of proton bunch being detached from the deuterium layer, reducing further acceleration of deuterium ions. 
	
\begin{figure}[ht]
			\begin{center}
				\includegraphics[width=\linewidth]{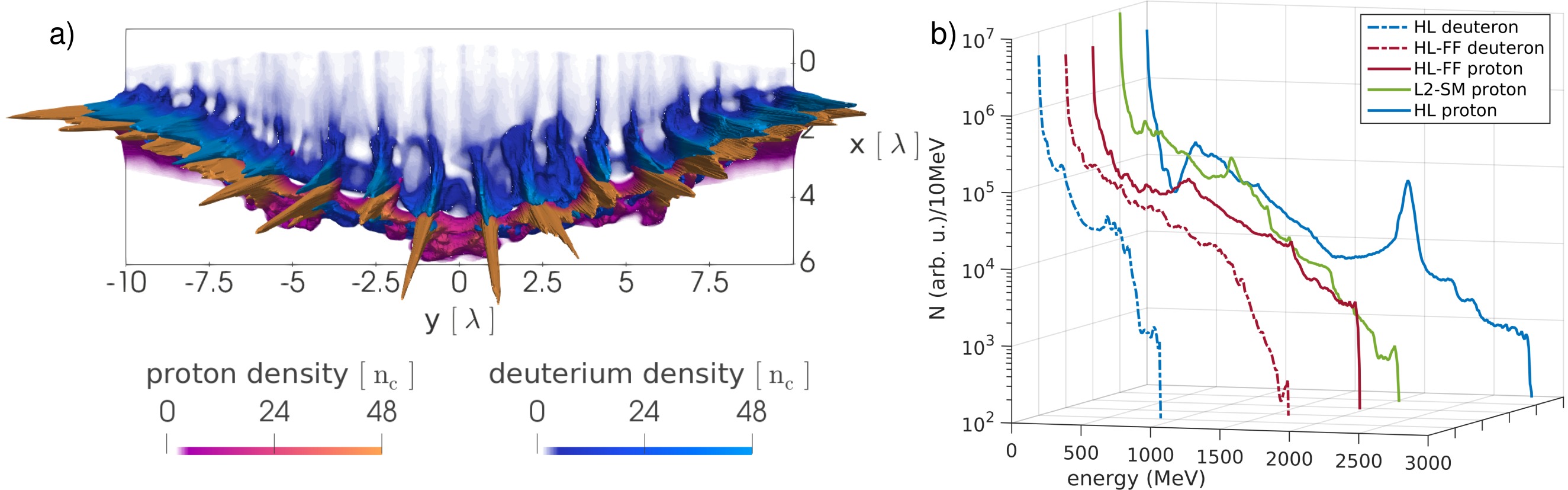} 
				\caption{\label{fig:SWI}Short-wavelength instability. a) Spatial distributions in the HL-FF case at time $ t = 14\ T $. Blue and red scales represent deuterium and proton densities with maximum value set to the initial density. Full density is indicated by the vertical height. b) Proton and deuteron energy spectra for HL, HL-FF and L2-SM cases at time $ t = 47\ T$.}
			\end{center}
		\end{figure}

	The development of the transverse SWI is shown in Fig. \ref{fig:SWI}-a. Here the full-front Gaussian laser pulse (8 $ T $ longer) is used instead of the pulse with the steep front. The rest of the simulation parameters, including the corrugation, are same as in the previous HL case. This simulation case is referred as HL-FF (full front).  The foil is disrupted into relatively high number of small bunches and low-density regions (compared to 3 large bunches at the same time instant in Fig. \ref{fig:HustEn}-a).
	
	On the first hand, the foil disruption enhances the acceleration of deuterons  as they are more mixed with the protons and are kept together for longer time. Deuterons are then accelerated to maximum energy more comparable with protons  (1600 MeV and 1920 MeV for deuterons and protons, respectively) than in the case of the steep-front pulse (870 MeV and 2770 MeV) as can be seen in Fig. \ref{fig:SWI}-b at the reference time $47\ T $. The data for the HL-FF case are shown at time postponed by $8\ T $, i.e., the peak intensity of both laser pulses reaches the target at the same time instant. 
	
	On the other hand, the foil disruption significantly reduces the potential of the radiation pressure acceleration of protons. In this case the maximal proton energy is reduced to 1920 MeV and only a relatively small peak in medium energies (with average energy of 690 MeV) is created (Fig. \ref{fig:SWI}-b). This behavior corresponds to previously made simulations with full pulse and single layer targets \cite{Pegoraro2007}.  Moreover, the proton spectrum is also similar to another simulation of the steep-front laser pulse (the same as in
	previous HL and LH cases) with pure 2 $ \lambda $ thick hydrogen
	target with the same modulation introduced on its front
	surface instead of the interface (referred as L2-SM). Where the average energy of the peak is 800 MeV and the maximum energy reaches 2010 MeV (Fig. \ref{fig:SWI}-b). 
			
\section{Conclusions}\label{Conclusion}

In conclusion, collisionless relativistic instability with RMI-like features is observed in our simulations, using double-layer targets with initial interface modulation and high-intensity steep-front laser pulse. Evolution of this instability is described in the heavy-light (HL) and light-heavy (LH) cases, resulting into a development of bunch structures in the density distribution and proton spectra. 

Well-collimated, quasi-monoenergetic proton bunch is observed for the HL case, with the average energy in the multi-GeV range, energy spread down to 3.7 \%, solid angle of $ 0.1\ \mathrm{mrad} $, divergence of 0.038 rad and transverse rms emittance down to 0.046 $\mathrm{m m\cdot mrad}$ (at FWHM). The emittance is one order of magnitude
lower compared to that of conventional accelerators. The energy conversion of the laser pulse into the proton particles inside this bunch is around 3.4 \% and overall conversion into all particles is over 40 \%.

Optimal parameters for the ultra-low energy spread are acquired using corrugation wavelength of half of the laser focal spot diameter and linear (s- in the 2D geometry) polarisation. 

The bunch is more distinctive in the proton spectrum and its average energy is
significantly higher compared to other simulated cases with the same target composition. These other cases are either dominated by  essentially different type of instabilities (development of short-wavelength instability using full-front laser pulse or introduction of the modulation on the surface of a single layer target) or the instability and bunch structures are not developed (targets without initial modulation using steep-front laser pulse, reducing the development of short-wavelength instability).

\appendix		
	\section{Data visualisation and virtual reality}\label{Appendix}
\begin{figure}[ht]
		\begin{center}
			\includegraphics[width=\linewidth]{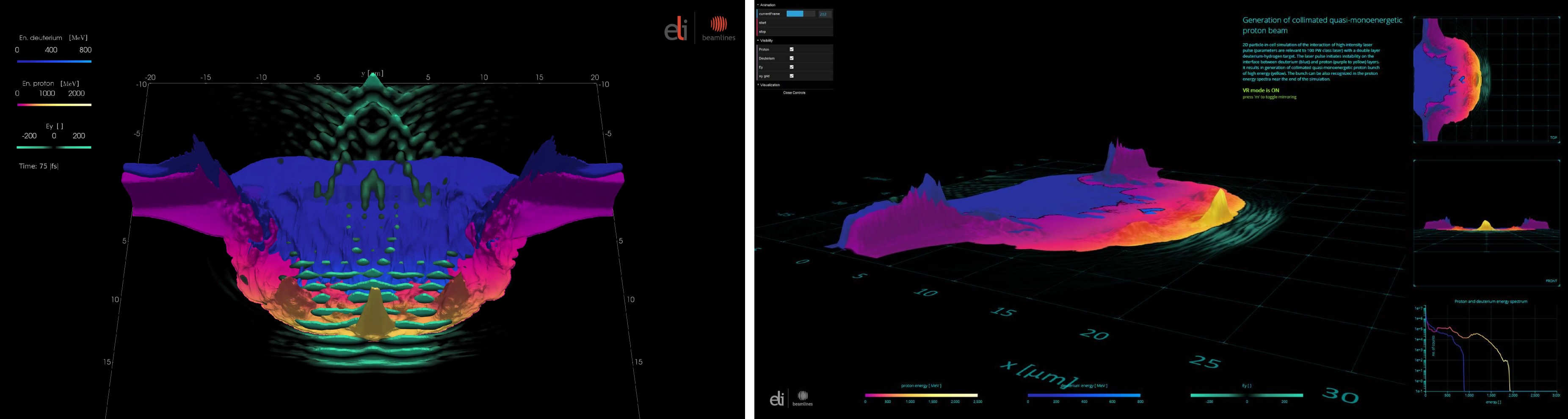}
			\caption{\label{fig:VR}Data visualisation.  Time evolution of the simulated HL case visualised in: a) ParaView, b) web-based interactive 3D application.}
		\end{center}
	\end{figure}

	Several visualization methods were used to create visual outputs from the above discussed simulation data. Firstly, raw data were imported in ParaView \cite{Ahrens2005} software to interpret them and to create visual outputs (figures and videos) describing the simulation mechanisms (see Fig. \ref{fig:VR}-a). Secondly, data have been presented in web-based interactive 3D application \cite{VBL} (see Fig. \ref{fig:VR}-b), which runs in a regular web browser and utilizes VR (virtual reality) mode to offer scientists a completely new point of view of their simulations. 
	
Here we present the videos (accessible online) of the time evolution of the simulated HL case mentioned above and in Fig. \ref{fig:VR} visualised in ParaView:

\begin{description}\label{VID1}
	\item[VID1.mp4] 
\end{description}

and in the web-based interactive 3D application: 

\begin{description}\label{VID1}
	\item[VID2.mp4]
\end{description}	
	The second option comes together with a challenge to find the best workflow that would enable visualization of such a large datasets in a web browser while maintaining high frame rates for smooth experience not only while using the VR headset. We have evaluated existing technologies like VTK.js library developed directly for scientific visualization in browser as well as Three.js javascript 3D library. Unfortunately, both libraries could not render our datasets with hundreds of timeframes with sufficient performance. The main issues were minutes-long loading times together with insufficient frame rates for large animated datasets. Thus, we have developed custom WebGL solution \cite{VBL}, a framework that not only renders the dataset on the GPU in real-time at high frame rates, but also provides orthogonal views, textual and numeric information, alongside a graphical user interface containing timeline animation controls and layer visibility management, with additional graphical elements based on D3.js for plotting animated graphs and legends. In order to import simulation data to this application, they are transformed into binary buffers in a node.js script so the visualization engine can directly send them to GPU with no need for further pre-processing. Currently we are working on an implementation of our transformation pipeline to the Jupyter notebook \cite{Jupyter} toolchain to allow more users to create high-performance web-based VR-enabled visualizations.
		
	\section*{Acknowledgements}
	
	Our work was supported by projects High Field Initiative (CZ.02.1.01/0.0/0.0/15\_003/0000449) and Extreme Light Infrastructure Tools for Advanced Simulation (CZ.02.1.01/0.0/0.0/16\_013/0001793) from the European Regional Development Fund, by Czech Science Foundation project 18-09560S and by the Grant Agency of the Czech Technical University in Prague, grant No. SGS19/192/OHK4/3T/14. This work was supported by The Ministry of Education, Youth and Sports from the Large Infrastructures for Research, Experimental Development and Innovations project "IT4Innovations National Supercomputing Center -- LM2015070", which provided computer resources for simulations. Fruitful discussions with Professor O. Klimo from FNSPE, CTU in Prague and ELI Beamlines (FZU - IoP CAS) are gratefully acknowledged. Collaboration with our colleagues from the Virtual Beamline team at ELI Beamlines including development of web-based interactive 3D application is gratefully acknowledged.

	\section*{References}
	
	\bibliography{references}

\end{document}